\documentclass[
 aip,
 amsmath,amssymb,
 reprint,
 longbibliography
]{revtex4-1}

\usepackage{graphicx}
\usepackage{dcolumn}
\usepackage{bm}
\usepackage{color}

\usepackage[utf8]{inputenc}
\usepackage[T1]{fontenc}
\usepackage{mathptmx}
\usepackage{natbib}

\begin{document}

\preprint{AIP/123-QED}

\title{Long Phonon Mean Free Paths Observed in Cross-plane Thermal-Conductivity Measurements of Exfoliated Hexagonal Boron Nitride}
 
 
\author{Gabriel R. Jaffe}
\email{gjaffe@wisc.edu}
\author{Keenan J. Smith}
\affiliation{Department of Physics, University of Wisconsin-Madison, Madison, Wisconsin 53706, USA}

\author{Kenji Watanabe}
\affiliation{Research Center for Functional Materials, National Institute for Materials Science, 1-1 Namiki, Tsukuba 305-0044, Japan}
\author{Takashi Taniguchi}
\affiliation{International Center for Materials Nanoarchitectonics, National Institute for Materials Science,  1-1 Namiki, Tsukuba 305-0044, Japan}

\author{Max G. Lagally}
\affiliation{Department of Materials Science and Engineering, University of Wisconsin-Madison, Madison, Wisconsin 53706, USA}

\author{Mark A. Eriksson}
\affiliation{Department of Physics, University of Wisconsin-Madison, Madison, Wisconsin 53706, USA}

\author{Victor W. Brar}
\affiliation{Department of Physics, University of Wisconsin-Madison, Madison, Wisconsin 53706, USA}


\begin{abstract}
Sub-micron-thick layers of hexagonal boron nitride (hBN) exhibit high in-plane thermal conductivity, useful optical properties, and serve as dielectric encapsulation layers with low electrostatic inhomogeneity for graphene devices.  Despite the promising applications of hBN as a heat spreader, the thickness dependence of the cross-plane thermal conductivity is not known, and the cross-plane phonon mean free paths in hBN have not been measured.  We measure the cross-plane thermal conductivity of hBN flakes exfoliated from bulk crystals.  We find that the thermal conductivity is extremely sensitive to film thickness.  We measure a forty-fold increase in the cross-plane thermal conductivity between 7\,nm and 585\,nm flakes at 295\,K.  We attribute the large increase in thermal conductivity with increasing thickness to contributions from phonons with long mean free paths (MFPs), spanning many hundreds of nanometers in the thickest flakes.  When planar twist interfaces are introduced into the crystal by mechanically stacking multiple thin flakes, the cross-plane thermal conductivity of the stack is found to be a factor of seven below that of individual flakes with similar total thickness, thus providing strong evidence that phonon scattering at twist boundaries limits the maximum phonon MFPs.  These results have important implications for hBN integration in nanoelectronics and improve our understanding of thermal transport in two-dimensional materials.
\end{abstract}

\maketitle

Effectively dissipating heat away from hotpots caused by high-power or densely packed electronic structures is an outstanding thermal management problem. Heat spreading films must exhibit high thermal conductivity, good dielectric characteristics, and form smooth clean interfaces with heat-emitting structures in order to reduce hotspot temperatures. Hexagonal boron nitride (hBN), a wide-bandgap dielectric two-dimensional (2D) material, has drawn significant research interest for its high in-plane thermal conductivity,\cite{Sichel_PRB_1976, Jo_NanoLett_2013, Yuan_CommPhys_2019} its use as a charge-trap free encapsulation material for graphene electronics,\cite{Dean_NatNanotech_2010,Calado_NatNanotech_2015} and its optical characteristics.\cite{Cassabois_NatPhotonics_2016}  Films of hBN can be mechanically cleaved from bulk crystals and transferred with atomically smooth and clean interfaces.  Such transferred films have demonstrated significant heat spreading capability in LED devices and graphene electronics.\cite{Choi_AMI_2018,Choi_AMI_2019}  The clean and conformal nature of these flexible hBN films have distinct advantages over materials such as diamond or SiC, which require thermally resistive interface layers in order to bond to other materials and suffer from growth defects near the interface layer.\cite{Won_IEEE_2015,Gu_JEM_2021}

The rate at which a hotspot can be cooled is determined by the strength of the three dimensional heat flow through a surrounding heat spreading film. Many 2D materials exhibit a high degree of anisotropy in their thermal conductivities, however, with some of the highest known conductivities along their in-plane directions but relatively small cross-plane thermal conductivities. While  their high in-plane thermal conductivities make these materials excellent candidates for heat spreading in nanoelectronics,\cite{Yam_NatCom_2012, Fu_2DMat_2019} the anisotropic thermal transport properties have applications in thermoelectrics and thermal isolation of temperature sensitive components.\cite{Chen_NanoRes_2015,Kim_JEP_2020,Jeon_ASS_2021,Li_NatRev_2021} The thermal conductivities of most bulk 2D crystals are well known; however, at sub-micron length scales the thermal conductivity of 2D materials can exhibit a strong thickness dependence.  Once the thickness of a film is less than the average phonon mean free path, the thermal conductivity begins to decrease. It is therefore necessary to know the mean free paths (MFPs) of the phonons responsible for heat transport in order to form an accurate model of thermal conductivity in the sub-micron regime.\cite{Chen_JHT_1996}  Recently, the phonon MFP spectra of graphite and MoS$_2$ were measured, and the phonon MFPs were found to be hundreds of nanometers at room temperature, far exceeding kinetic theory estimates.\cite{Zhang_NanoLett_2016,Sood_NanoLett_2018} It is currently not known whether other two-dimensional materials also exhibit these long phonon MFPs, and to what degree scattering mechanisms such as grain boundaries limit the maximum phonon MFPs.

 The cross-plane thermal conductivity of bulk hBN ($>$10\,$\mu$m thick) has been measured,\cite{Simpson_JPCSSP_1971, Jiang_PRM_2018, Yuan_CommPhys_2019} and a first-principles calculation predicted the average cross-plane phonon MFP to be $\sim$18\,nm at 300\,K.\cite{Jiang_PRM_2018} This prediction is an order of magnitude lower than the average MFPs measured in graphite and MoS$_{2}$ and has not been verified experimentally. The average cross-plane phonon mean free path, $\Lambda_{avg}$, can be estimated empirically from kinetic theory using the formula $\kappa_{bulk} = (1/3)\,C\,v_{g}\,\Lambda_{avg}$.  For hBN the cross-plane thermal conductivity\cite{Simpson_JPCSSP_1971, Jiang_PRM_2018} $\kappa_{bulk} \sim$2--5\,W\,m$^{-1}$K$^{-1}$, heat capacity\cite{Simpson_JPCSSP_1971} $C \sim$1.8\,J\,cm$^{-3}$K$^{-1}$, and the average group velocity of the cross-plane acoustic phonon modes\cite{Jiang_PRM_2018} $v_{g} \sim$3,800\,m\,s$^{-1}$ yield a mean free path estimate of 0.9--2.2\,nm.
 
In this Article, we report thermal-conductivity measurements of exfoliated hBN flakes. We find that the cross-plane thermal conductivity  at 295\,K increases by more than a factor of 40 with flake thickness, from 0.20\,$\pm$\,0.06\,W\,m$^{-1}$K$^{-1}$ for a 7\,nm flake to 8.1\,$\pm$\,0.5\,W\,m$^{-1}$K$^{-1}$ for a 585\,nm flake.  Fits to the data indicate that the majority of the heat is carried by phonons with mean free paths $>$100\,nm. This value far exceeds the MFPs estimated from kinetic theory, which predicts MFPs of only a few nanometers, as well as the prediction from first-principles calculations that $\sim$80\% of the heat is carried by phonons with MFPs between 3--90\,nm.\cite{Yuan_CommPhys_2019}  Furthermore, we find that when planar twist boundaries are intentionally introduced by mechanically stacking five exfoliated flakes, the stacked structure is found to have a cross-plane thermal conductivity a factor of seven below that of individual flakes with the same total thickness as the stack.  We attribute the low thermal conductivity of the stacked hBN flakes to a suppression of the phonon MFPs through phonon scattering at twist interfaces.

\begin{figure}[t]
	\centering
	\includegraphics[width = \columnwidth]{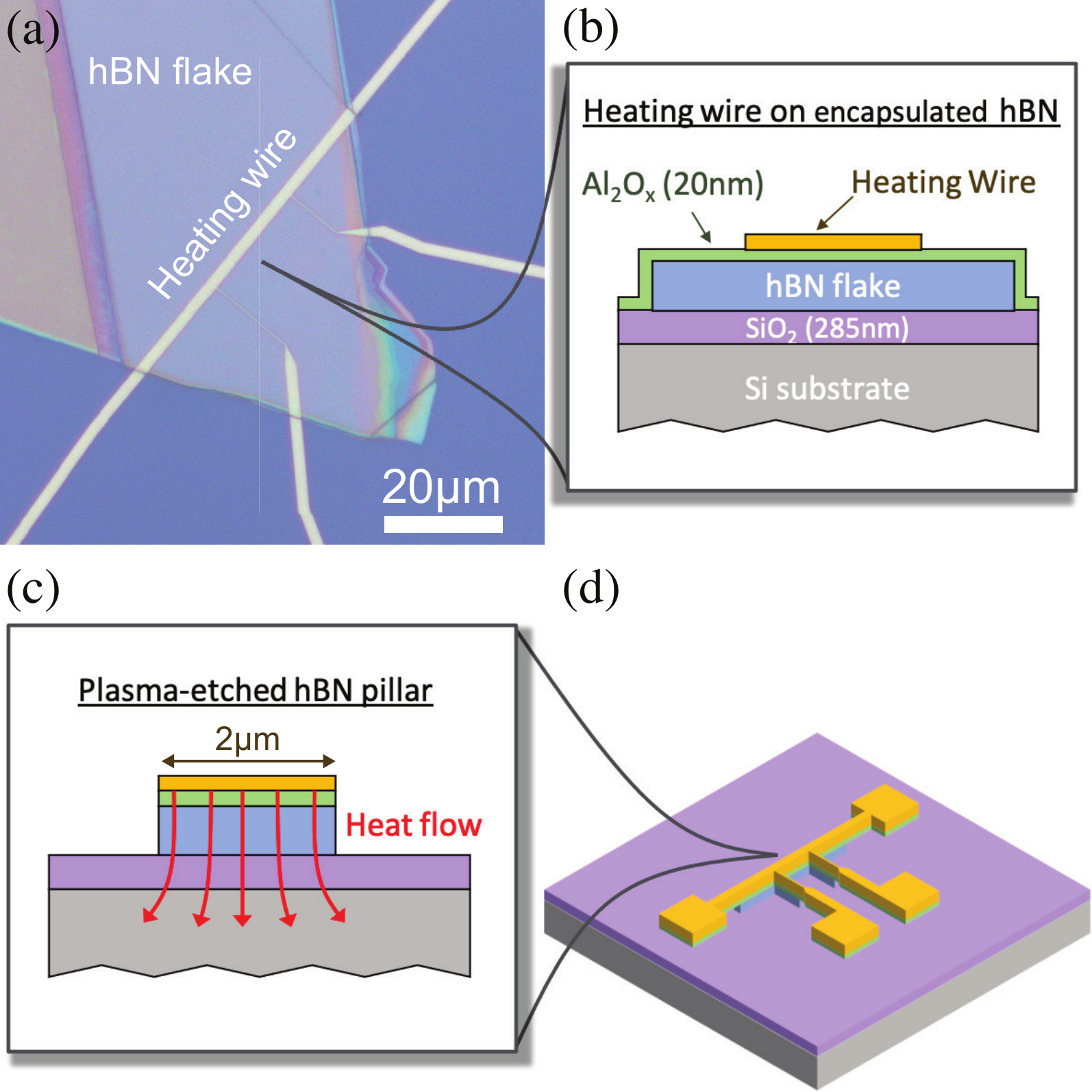}
	\caption{(a) An optical image of a heating wire fabricated across an AlO$_x$-encapsulated hBN flake that has been exfoliated from a bulk crystal and mechanically transferred to a SiO$_2$/Si substrate. (b)  A cross-sectional view of the heating wire on the flake. (c) A cross-sectional view showing how the encapsulating AlO$_x$ layer and underlying hBN are etched into a pillar after BCl$_3$Ar and SF$_6$ plasma etches, respectively.  During a measurement, the heat from the wire flows cross-plane through the hBN and dissipates into the substrate.  (d) An isometric view of the completed four-probe heater/thermometer wire on hBN.  The scale is exaggerated for clarity.}
	\label{fig:1}
\end{figure}

Taken together, the long MFPs and strong interface scattering observed here have two important implications.  First, the long phonon MFPs in hBN impact its use in heat spreading applications in nanoelectronics, because the cross-plane thermal conductivity of hBN will decrease dramatically as the film thickness is reduced below the phonon MFPs.  Heat dissipation from hotspots in electronics flows both in-plane and cross-plane through a heat-spreading film.  It is often desirable to make heat-spreading films as thin as possible so as to maximize their cross-plane thermal conductance.  Our results demonstrate that the cross-plane thermal conductance of hBN does not significantly increase for films $<$\,300\,nm thick. Second, these results demonstrate that two-dimensional materials are a promising model system for studying coherent phonon transport behavior, because heterostructures of dissimilar materials with layer thicknesses far below the typical phonon MFPs can be fabricated with smooth and pristine interfaces by stacking different exfoliated flakes at arbitrary angles.  

\begin{figure}[t]
	\centering
	\includegraphics[width = \columnwidth]{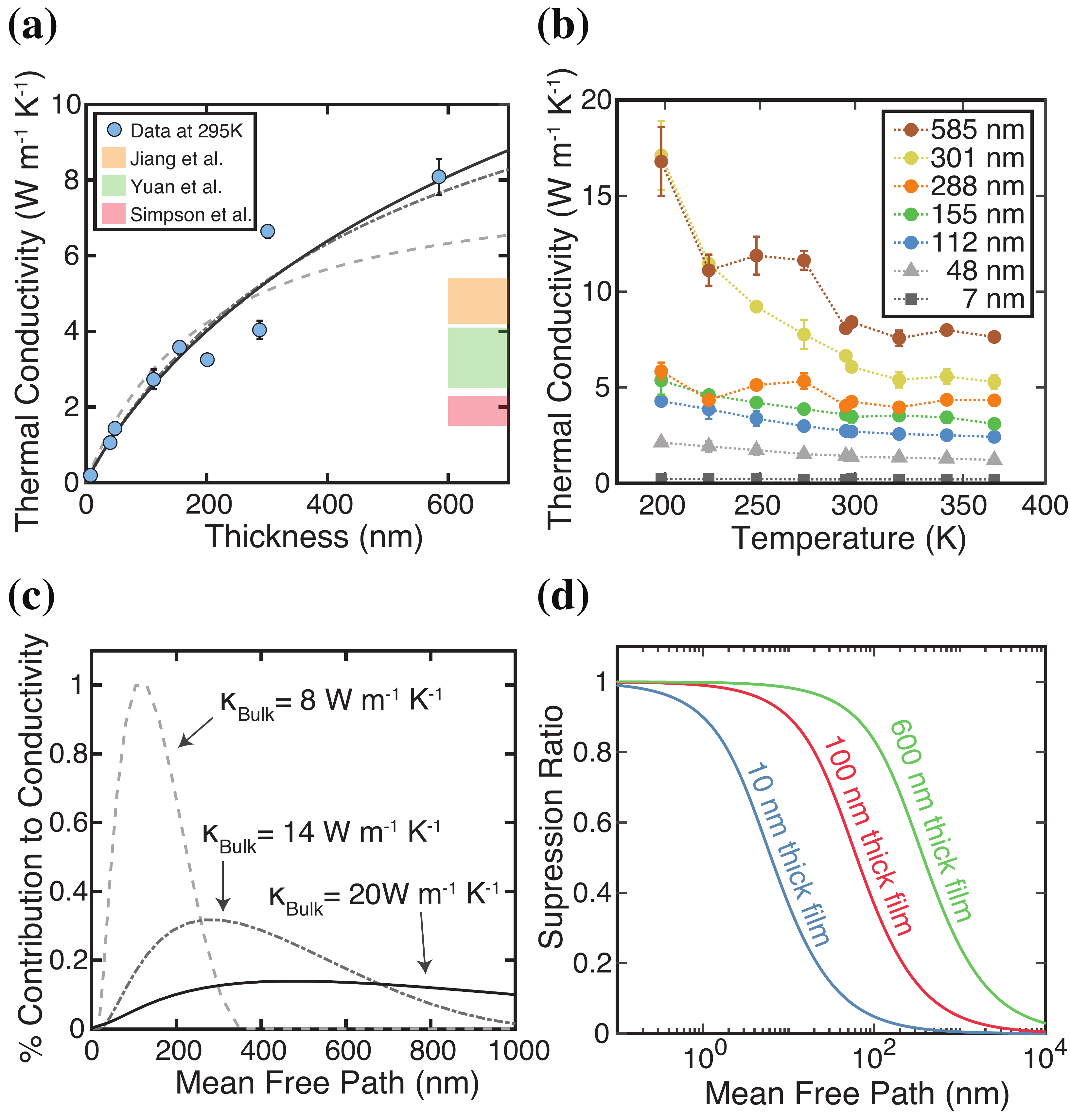}
	\caption{(a) The measured cross-plane thermal conductivity (blue points) of hBN flakes exfoliated from bulk crystals as a function of flake thickness at 295\,K.  The lines show the phonon mean free path distribution fits to the data assuming that the thermal conductivity eventually saturates to either 8, 14, or 20\,W\,m$^{-1}$K$^{-1}$. Shaded regions indicate the thermal conductivity of bulk (>10$\mu$m) hBN measured by others.\cite{Simpson_JPCSSP_1971, Jiang_PRM_2018, Yuan_CommPhys_2019}  (b) The cross-plane thermal conductivity as a function of temperature of the hBN flakes with colors indicating the flake thicknesses.  The dashed lines serve to help guide the eye. (c) The differential phonon man free path spectra that best fit the data shown in (a) assuming three different values for the bulk thermal conductivity $\kappa_{Bulk}$.  (d)  The suppression of contributions to thermal conductivity (Eq.\,\ref{eq:1}) of phonons with long mean free paths for films of three different thicknesses.}
	\label{fig:2}
\end{figure}

We measure the cross-plane thermal conductivity of hBN flakes using the three-omega method.\cite{Cahill_RSI_1990}  For this technique, a metal four-probe wire is fabricated across the film of interest and is then used as both a heater and thermometer.  These measurements are performed in a differential configuration, with the signal from a heating wire placed directly on the substrate subtracted from the signal measured with a heating wire on a pillar of hBN.\cite{Cahill_PRB_1994, Tasciuc_SM_2000, Koh_JAP_2009, Jaffe_ACSAMI_2019, Jaffe_APL_2020}  Fig.\,\ref{fig:1}(a) shows an optical image of a heating wire on an hBN flake.  For all samples, the heating wire is 2\,$\mu$m wide, the distance between the voltage probes is 20\,$\mu$m, and the voltage probe widths are 300\,nm where they make contact with the heating wire.  Bond pads are fabricated $>$\,400\,$\mu$m away from the section of wire between the voltage probes in order to prevent heat dissipating out through the wire bonds from affecting the experiment.  

The hBN flakes are exfoliated in a glove box with nitrogen atmosphere and transferred onto SiO$_{2}$/Si substrates.  The flakes are vacuum annealed under an Ar/H$_2$ flow at 350\,$^{\circ}$C for one hour.  An atomic force microscope is used to measure the flake thicknesses and surface roughnesses.  The flakes are encapsulated in a 20\,nm thick layer of AlO$_x$ deposited by atomic-layer deposition to protect the interface between the flake and substrate from solvent contamination during processing.  Metal four-probe heater/thermometer wires are fabricated on the sample surface using electron-beam lithography. The Au wires are 65\,nm thick with a 5\,nm Ti adhesion layer.  Small metal interconnects are patterned on the edges of flakes with thicknesses $>$\,60\,nm to connect the wires over the large jumps in the surface topography.  In order to ensure that the heat flowing out of the wire through the hBN flakes is entirely cross-plane, the hBN flakes are etched into pillars, as shown in Fig.\,\ref{fig:1}(c), using first a BCl$_3$Ar plasma etch to remove the AlO$_x$ layer, and then a SF$_6$ plasma etch to remove the hBN.

Figure\,\ref{fig:2}(a) shows the measured cross-plane thermal conductivity of the hBN flakes as a function of flake thickness at 295\,K.  We observe that the thermal conductivity more than doubles as the flake thickness increases from 200\,nm to 585\,nm, indicating that phonons with MFPs of several hundred nanometers make significant contributions to the thermal conductivity of the thicker flakes.  Moreover, the thermal conductivity appears to be gradually saturating with increasing thickness, consistent with similar trends seen in graphite.\cite{Zhang_NanoLett_2016}  For the thickest flakes that we measure, the thermal conductivities exceed previously measured bulk values, which are shown in Fig.\,\ref{fig:2}(a) as shaded orange, green, and red squares.\cite{Simpson_JPCSSP_1971, Jiang_PRM_2018, Yuan_CommPhys_2019}  We note that the thermal conductivity of hBN is known to vary significantly with crystal defect density, and that hBN samples with the lowest previously measured cross-plane thermal conductivity ($\sim$2\,W\,m$^{-1}$\,K$^{-1}$) were found to have an average crystallite size in the cross-plane direction of $\sim$10\,nm.\cite{Simpson_JPCSSP_1971}

Fig.\,\ref{fig:2}(b) shows the cross-plane thermal conductivity of flakes of different thicknesses as a function of temperature.  The thermal conductivity increases with decreasing temperature, as is expected in ordered materials where Umklapp scattering dominates.  As the temperature decreases, the average wavelength of the excited phonons shifts to longer wavelengths, yet phonons with wavelengths longer than the crystal dimensions cannot be formed.  It is therefore expected that thinner flakes will display less temperature dependence in their thermal conductivities than thicker flakes, as is observed.  The thermal interface resistances at the top and bottom interfaces of the flakes, which do not vary significantly with temperature, could also be contributing to the weak temperature dependence of the thinnest flakes. 

For an infinitely thick film, the differential mean free path contribution function $f(\Lambda)$ describes the fractional contribution of phonons with mean free path $\Lambda$ to the thermal conductivity.\cite{Minnich_PRL_2012}  The contributions from phonons with long mean free paths are suppressed in a film of finite thickness.  The suppression of phonon contributions to thermal conductivity for the sample geometry considered here is described by the suppression function
\begin{flalign}\label{eq:1}
	& \quad S(K_n) = 1 - K_n (1-e^{-\frac{1}{K_n}}), &
\end{flalign}
where the Knudsen number $K_n$ is defined as $K_n = \frac{\Lambda}{L}$ and $L$ is the film thickness.\cite{Zhang_NanoLett_2016}  The thermal conductivity of a thin film as a function of film thickness $\kappa(L)$ is a convolution of $f(\Lambda)$ and $S(K_n)$ over all phonon mean free paths
\begin{flalign}\label{eq:2}
	& \quad \kappa(L) = \int_0^{\infty} S(K_n) f(\Lambda)\,d\Lambda.  &
\end{flalign}
We find the differential mean free path contribution functions that best fit our thermal conductivity data at 295\,K using a convex optimization procedure and a Gaussian quadrature discretization of the integral in Eq.\,\ref{eq:2}.\cite{Minnich_PRL_2012}  Additional information about the fitting procedure is available in the supporting information.  Although the thermal conductivity appears to be rolling off at large thicknesses in Fig.\,\ref{fig:2}(a), we do not know the bulk thermal conductivity of an infinitely thick flake.  We therefore provide --- in Fig.\,\ref{fig:2}(a) --- the mean free path fits to the data assuming a bulk conductivity of either 8, 14, or 20 W\,m$^{-1}$K$^{-1}$.  The best-fit phonon mean free path contribution functions are shown in Fig.\,\ref{fig:2}(c).  In all fits we find that phonons with mean free paths $>$\,100\,nm are responsible for the majority of the thermal transport.  The thermal conductivities as a function of thickness calculated from the fits are shown as lines in Fig.\,\ref{fig:2}(a).  We find that all the fits are in close agreement for thicknesses $<$\,200\,nm and begin to diverge at larger thicknesses.

\begin{figure}[t]
	\centering
	\includegraphics[width= \columnwidth]{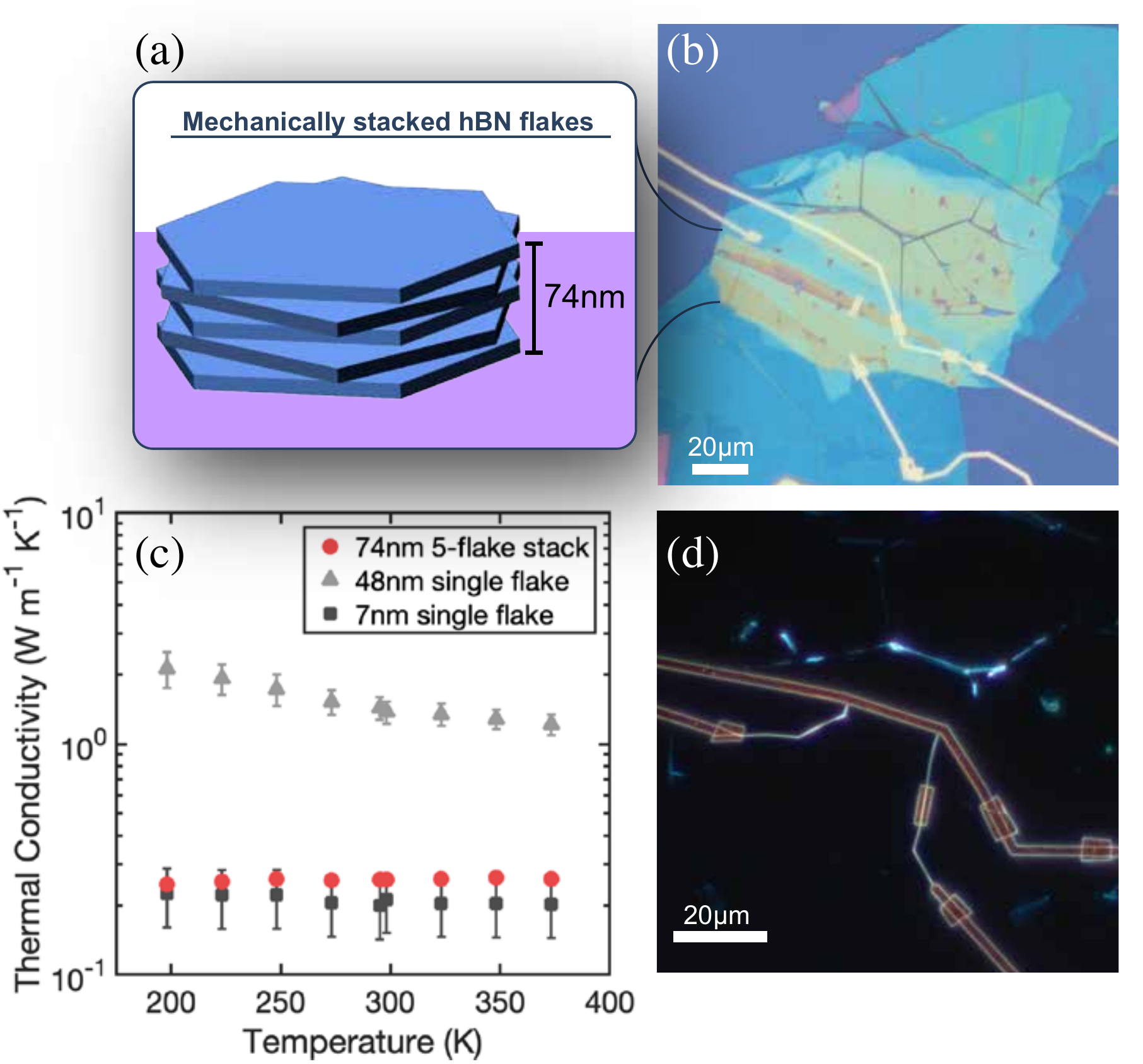}
	\caption{(a)  A schematic diagram showing a heterostructure consisting of five exfoliated hBN flakes mechanically stacked with arbitrary rotational mismatches between each flake.  (b) An optical image of the heating wire fabricated across the five-flake stack.  (c) The measured cross-plane thermal conductivity as a function of temperature of the five-flake stack shown in (b), which had a total thickness of 74nm, is shown in red.  Grey squares and triangles are the thermal conductivities of individual flakes from Fig.\,\ref{fig:2}(b) with thicknesses of 7nm and 48nm, respectively. (d)  A dark-field optical image of the five-layer hBN stack sample after plasma etching the hBN into a pillar.}
	\label{fig:3}
\end{figure}

Phonon grain boundary scattering at planar twist interfaces has been proposed as a limiting factor for the maximum possible phonon MFPs in two-dimensional materials.  Studies of graphite correlated the maximum observed phonon MFPs with the average spacing between twist grain boundaries found in cross-sectional TEM images.\cite{Zhang_NanoLett_2016}    This hypothesis is further supported by studies of WSe$_{2}$ crystals grown with random rotational mismatches between each successive layer that found that twist interfaces reduced the cross-plane thermal conductivity by a factor of 30 below that of single-crystal samples.\cite{Chiritescu_SCI_2007}  Molecular-dynamics simulations of twist interfaces of both graphite and hBN have also shown that such grain boundaries can significantly reduce the cross-plane thermal conductivity.\cite{Ouyang_NanoLett_2020}

In order to investigate the effects of grain boundary scattering in hBN, we introduce twist interfaces into a hBN crystal by mechanically stacking five exfoliated hBN flakes.  The total stack thickness is measured with an atomic force microscope to be 74\,$\pm$\,2\,nm, and the individual layer thicknesses are estimated using optical contrast to be (in order of bottom to top) 26, 19, 11, 10, 8\,nm, respectively.  Fig.\,\ref{fig:3}(a) shows a schematic diagram of the stacked flakes, and Fig.\,\ref{fig:3}(b) shows an optical image of the heating wire fabricated over the stack before the hBN is etched into a pillar.   The dark-field optical image in Fig.\,\ref{fig:3}(d) was taken after etching the hBN and shows no evidence of residue or wrinkles near the heating wire.  

We measure the cross-plane thermal conductivity of the five-flake stack and compare the results to the thermal conductivity of individual flakes of similar thicknesses as a function of temperature in Fig.\,\ref{fig:3}(c).  We calculate the expected thermal conductivity of a 74\,nm flake to be 2.00\,W\,m$^{-1}$K$^{-1}$ at 295\,K by averaging the values from the fits at that thickness in Fig.\,\ref{fig:2}(a).  By comparison, we measure the thermal conductivity of the five-flake stack to be 0.26\,$\pm$\,0.01\,W\,m$^{-1}$K$^{-1}$, more than a factor of 7 below this estimate.  This indicates that strong phonon scattering at the twist interfaces has drastically reduced the phonon MFPs, thereby reducing the thermal conductivity.  Furthermore, the thermal conductivity of the stack does not increase at lower temperatures, suggesting that the thermal conductivity is dominated by interface scattering.  

Our results demonstrate that the cross-plane thermal conductivity of hBN is highly tunable.  We see a factor of 40 increase in the thermal conductivity as the film thickness is increased from 7 to 585\,nm.  Fits to the data reveal that the heat is being transported by phonons with MFPs hundreds of nanometers in length, far exceeding earlier theoretical predictions.  We demonstrate that the thermal conductivity of thicker films of hBN can be significantly reduced by stacking multiple thin flakes of hBN with arbitrary rotational mismatches between each layer. The ability to stack hBN flakes at controlled angles and potentially incorporate layers of other two-dimensional materials with long phonon MFPs, such as graphite and MoS$_{2}$, presents interesting opportunities to explore phonon interface scattering as parameters such as twist angle, layer thickness, and phonon mode mismatch between layers are varied.  The data presented here have important implications for thermal-management efforts to incorporate hBN as a heat-spreading material, providing the necessary information for determining how the cross-plane thermal conductivity of hBN films will scale with film thickness.

Thermal transport measurements and theoretical modeling were performed with support from US DOE Basic Energy Sciences DE-FG02-03ER46028.   hBN crystal exfoliation and stacking was performed with support from US DOE Basic Energy Sciences under Award \#DE-SC0020313.  We acknowledge the use of facilities supported by NSF through the UW-Madison MRSEC (DMR-1720415) and electron beam lithography supported by the NSF MRI program (DMR-1625348).  K.W. and T.T. acknowledge support from the Elemental Strategy Initiative conducted by the MEXT, Japan, grant number JPMXP0112101001,  JSPS KAKENHI grant number JP20H00354 and the CREST(JPMJCR15F3), JST.  The authors thank Dr.\,Donald Savage for his helpful comments and suggestions.


%

\end{document}